\documentclass{article}
\usepackage{spconf,amsmath,graphicx,hyperref}
\usepackage{amsmath,amsfonts}
\usepackage{mathtools,nccmath}
\usepackage{algorithmic}
\usepackage{algorithm}
\usepackage{array}
\usepackage{graphicx}
\usepackage{textcomp}
\graphicspath{{figures/}}
\usepackage{booktabs}
\usepackage{makecell}
\usepackage{indentfirst}
\usepackage{enumitem}
\usepackage{etoolbox}
\usepackage{multirow}
\usepackage{booktabs}
\usepackage{bm}

\makeatletter
\patchcmd{\thebibliography}
  {\small}
  {\footnotesize}
  {}{}

\patchcmd{\thebibliography}
  {\itemsep\z@}
  {\itemsep\z@\parskip\z@}
  {}{}
\makeatother

\title{Monotonicity-Guided Semantic Alignment for \\ Zero-shot Multispeaker Image-to-Speech Synthesis}
\name{Lijun Wang, Yixian Lu, Shogo Okada\textsuperscript{*}\thanks{* Corresponding Author. Email: okada-s@jaist.ac.jp.}}
\address{School of Information Science, Japan Advanced Institute of Science and Technology}
\begin{document}
%
\maketitle
\begin{abstract}
Direct image-to-speech (Img2Sp) poses an alignment challenge in mapping visual content to ordered speech sequences, as images permit multiple spoken descriptions and lack monotonic correspondence with speech sequences. We propose Monotonicity-Guided Semantic Alignment (MGSA), to the best of our knowledge, the first framework for zero-shot multispeaker Img2Sp synthesis. We use semantic speech units to provide shared content targets across speakers with reference speech for speaker conditioning. A query aligner maps semantic memory learned from visual content to speech unit positions via a soft monotonic prior, which yields position-specific conditioning states. A blockwise masked diffusion generator is employed for the speech unit generation conditioning on these states. Experiments on Flickr8k-Audio show competitive captioning performance against single-speaker baselines, while evaluation with LibriTTS-R references supports zero-shot synthesis for unseen speakers. Ablations validate the effectiveness of aligner and block diffusion. Audio samples are available at \url{https://alizeded.github.io/mgsa-demo}.
\end{abstract}

\section{Introduction}
\vspace{-0.3em}
\label{sec:intro}
Direct image-to-speech (Img2Sp) synthesis generates spoken descriptions directly from visual content without intermediate text, which is important when written form is unavailable or latency is critical, such as in hands- and eyes-free interaction and in-cabin assistive narration. It also expands speech technology to roughly half of the world's languages that lack standardized written forms \cite{lee-etal-2022-textless}. 

Some existing methods predict mel spectrograms \cite{wang2021show} or discrete unit representations \cite{hsu2021sat,effendi2021end} by using encoder-decoder architectures trained with teacher forcing. Im2Sp\,\cite{kim2024towards} transfers a pretrained image captioning model to autoregressive (AR) prediction of clustered self-supervised HuBERT features. PixVoxLM\,\cite{f2p2025} predicts EnCodec tokens \cite{defossez2022encodec} by fine-tuning BLIP \cite{li2022blip}. However, these methods do not demonstrate reference-conditioned synthesis (i.e., zero-shot multispeaker setting) for unseen speakers. For example, SAS and PixVoxLM use synthesized speech from a single-speaker text-to-speech (TTS) system \cite{wang2021show, f2p2025}. Im2Sp reconstructs the predicted units using a vocoder trained on LJSpeech \cite{kim2024towards}. We thus explore zero-shot multispeaker Img2Sp synthesis, where the system generates image-grounded descriptions while conditioning on reference speech from unseen speakers. 

To support the zero-shot multispeaker synthesis, we employ semantic speech units as shared content targets across speakers and reference speech for speaker conditioning. This choice is also motivated by the concern that regressing continuous features (e.g., mel spectrograms) may prevent the model from focusing on image content \cite{kim2024towards}. However, these units do not specify how visual content corresponds to speech unit positions. Compared with text, an image permits multiple descriptions with different content and ordering. Its spatial features do not have predefined monotonic correspondence to unit positions. Speaker-dependent variations in speaking rates and acoustic conditions further complicate alignment learning in open-set multispeaker settings, as observed in zero-shot TTS \cite{chen2020multispeech}. Reference speech and teacher-forcing prefixes also introduce predictive cues that may create a shortcut to reduce the model's reliance on image conditions, since neural networks can favor the most exploitable signals and underfit weaker ones \cite{geirhos2020shortcut,pezeshki2021gradient, wu2022characterizing}. 

Accordingly, the remaining challenge is to organize visual content into position-specific conditions for speech unit generation. We therefore introduce an alignment framework that uses a soft monotonic prior to guide encoded visual content to construct position-specific conditions for speech unit generation. An auxiliary objective is introduced to supervise the content at each speech unit position. The prior acts on the learned memory rather than the spatial order of image features. We adopt blockwise masked diffusion to combine flexible length generation and efficient blockwise inference \cite{arriola2025block}. Unlike non-AR approaches that require a predefined duration, it generates block-by-block until termination. Compared with token-by-token AR decoding, masking mechanism within each block limits access to local prefix context, potentially reducing the tendency to favor prefix continuation over image conditioning \cite{geirhos2020shortcut,pezeshki2021gradient}.

\begin{figure}
    \centering
    \includegraphics[width=1.0\linewidth]{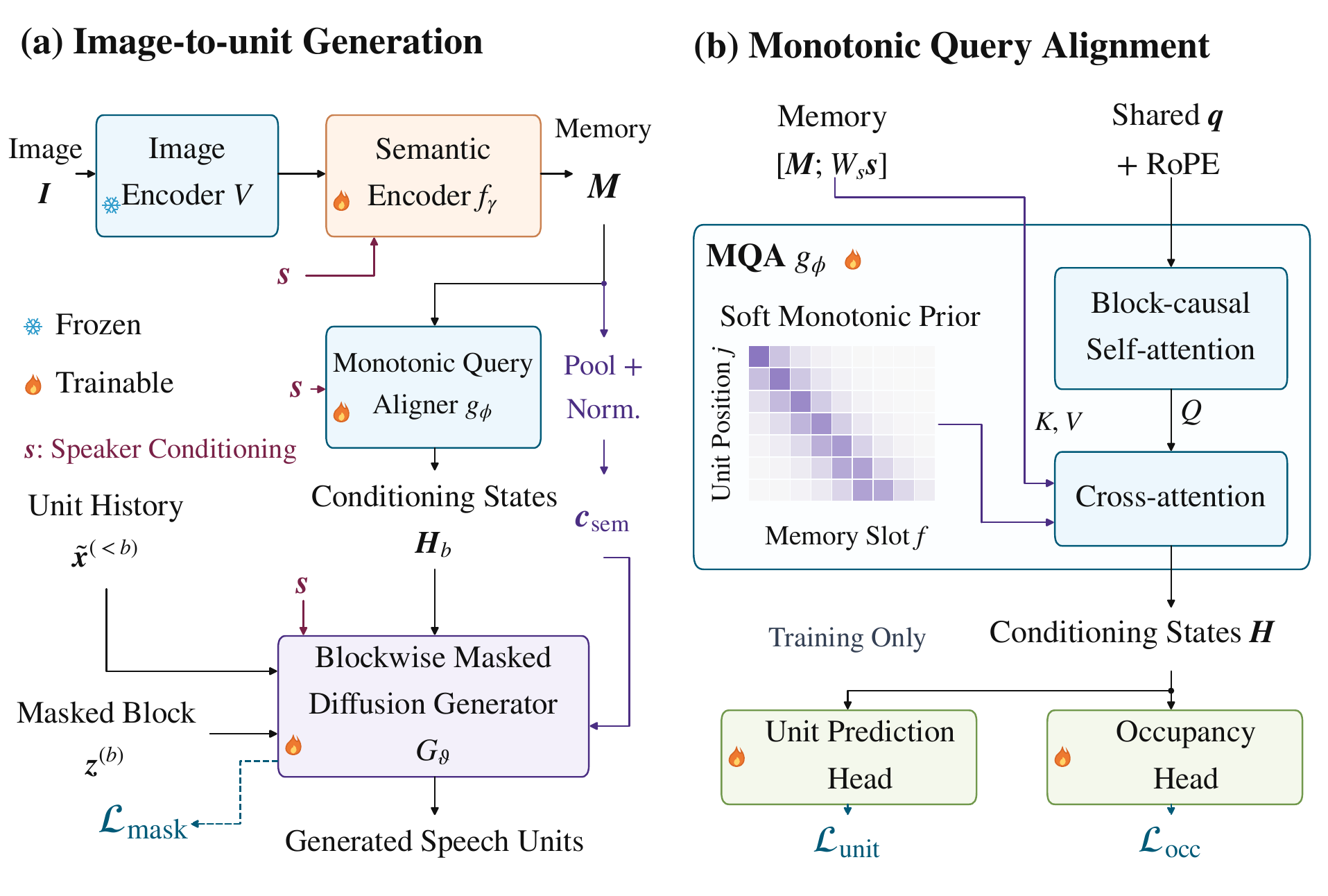}
    \vspace{-2.0em}
    \caption{Overview of the MGSA: A zero-shot multispeaker Img2Sp generation framework.}
    \label{fig:overview}
    \vspace{-1.3em}
\end{figure}

We propose Monotonicity-Guided Semantic Alignment (MGSA), as illustrated in Fig.\,\ref{fig:overview}. A semantic encoder (SE) $f_{\gamma}$ learns an indexed semantic memory from visual features and speaker conditioning. A monotonic query aligner (MQA) $g_{\phi}$ maps this memory to speech unit positions using a soft monotonic prior to construct conditioning states, without access to speech unit prefixes. This design encourages the states to encode visual content without relying on preceding units. A blockwise masked diffusion generator $G_{\vartheta}$ uses the states to generate speech units. Following \cite{wang2021show,f2p2025}, we train the image-to-unit (I2U) models and use an off-the-shelf unit-to-speech (U2S) model for speech waveform synthesis.

Our contributions can be organized as follows:
\vspace{-0.5em}
\begin{enumerate}[noitemsep]
    \item To our knowledge, we are the first to introduce and investigate zero-shot multispeaker Img2Sp synthesis.
    \item We propose MGSA, a new approach based on blockwise masked diffusion for zero-shot multispeaker synthesis using monotonicity-guided alignment.
    \item Experiments show that MGSA achieves competitive performance with single-speaker baselines while supporting unseen speakers with excellent speech quality.
\end{enumerate}

\vspace{-1.1em}
\section{Proposed Method}
\label{sec:prob_form}
\vspace{-0.2em}
\textbf{Preliminaries.} Given an image $\boldsymbol{I}$, reference speech $a_{\mathrm{ref}}(t)$, and an optional prefix prompt $\boldsymbol{x}_{<P}$, our goal is to synthesize speech $\hat{y}(t)$. A speech tokenizer $\mathcal{E}_{\mathrm{tok}}(\cdot)$ converts target speech $s(t)$ into discrete speech units $\boldsymbol{x} = [x_0, \cdots, x_{L-1}] \in \mathcal{V}^{L}$. A reference-conditioned U2S model $\mathcal{D}_{\mathrm{tok}}(\cdot)$ reconstructs the waveform as $\hat{y}(t) = \mathcal{D}_{\mathrm{tok}}(\hat{\boldsymbol{x}};\boldsymbol{a}_{\mathrm{ref}})$. Let $B$ be the block size and $P = kB$ the prefix length for a $k$-block prompt, with $P=0$ for generation without a prompt. The speaker style $\boldsymbol{s}$ is extracted from $a_{\mathrm{ref}}(t)$. $\boldsymbol{I}$ is the content condition. $a_{\mathrm{ref}}(t)$ offers speaker conditioning and the image guides the content of synthesized speech.

\textbf{Semantic Content Encoding.} A frozen visual encoder $\boldsymbol{V}(\cdot)$ first extracts image representations. The semantic encoder $f_{\gamma}$ then encodes visual features and speaker conditioning $\boldsymbol{s}$ into a fixed-capacity memory $\boldsymbol{M} \in \mathbb{R}^{F \times d}$\ with $F$ slots:
\begin{equation}
    \boldsymbol{M}=f_{\gamma}(\boldsymbol{V}(\boldsymbol{I}), \boldsymbol{s}).
\end{equation}
$\boldsymbol{M}$ stores semantic content in indexed slots. To provide global semantic conditioning $c_{\mathrm{sem}}$ for $G_{\vartheta}$, we mean-pool and normalize $\boldsymbol{M}$ following the practice of \cite{lee2025dittotts}.

\textbf{Monotonic Query Alignment.} While $\boldsymbol{M}$ provides semantic content, its slots have no predefined correspondence to speech unit positions. We introduce MQA $g_{\phi}$ to produce position-specific conditioning states $\boldsymbol{H}$ by attending to $\boldsymbol{M}$ with a soft monotonic prior. $\boldsymbol{H}$ are further fed into the generator $G_{\vartheta}$ to generate the unit sequences block by block. This component plays a central role in aligning visual content with ordered speech unit sequences. 

When generating block $b$, $G_{\vartheta}$ predicts the current block conditioned on all completed blocks. The MQA $g_{\phi}$ produces conditioning states $\boldsymbol{H}_b$ for $N_b=(b+1)B$ positions using a shared learnable query token $\boldsymbol{q}$ across positions and positional information from positional encoding.
\begin{equation}
\vspace{-.5em}
\label{tqd}
\boldsymbol{H}_b=g_{\phi}\Big((\boldsymbol{q}_{b,j})_{j=0}^{N_b-1}, M, W_s\boldsymbol{s}\Big) \in \mathbb{R}^{N_b\times d}.
\end{equation}
Here, $0 \le j < N_b$
We append $d$-dimensional speaker token $W_s\boldsymbol{s}$ to $\boldsymbol{M}$ to form the cross-attention memory. For block $b$, the input sequences stack $N_b$ 
queries $\boldsymbol{q}_{b,j}=\boldsymbol{q}$ with positions distinguished by positional encoding.

In both training and inference, MQA $g_{\phi}$ takes only shared queries, positional information, and the image and speaker conditions as input. Speech units are fed exclusively to the generator $G_{\vartheta}$ that uses ground-truth units as teacher-forced prefixes during training and generates units block by block during inference. In training, $g_{\phi}$ computes all $L+B$ states in a single forward pass. In inference, it recomputes the states for each new block with $B$ additional queries. Although all states are recomputed, the states in completed blocks remain unchanged since their inputs are fixed, and the block-causal mask prevents attention to future blocks.


\textbf{Block diffusion generation.} The unit sequence $\boldsymbol{x}$ is padded with the $\langle\mathrm{EOS}\rangle$ (end-of-sequence) token to the nearest block boundary, following \cite{arriola2025block}: $\tilde{\boldsymbol{x}}=(x_0, x_1,\cdots,x_{L-1},\\ \langle\mathrm{EOS}\rangle^{\widetilde{L}-L})$, where $\widetilde{L}=B\lceil(L+1)/B\rceil$. 
Let $\tilde{\boldsymbol{x}}^{(b)}=\tilde{\boldsymbol{x}}_{bB:(b+1)B}$ denote block $b$. The generator $G_{\vartheta}$ predicts per-position unit distributions $p_{\vartheta}$ through softmax, while $p_{\vartheta}^{\mathrm{blk}}$ denotes the block distribution induced by iterative masked decoding. For a prefix of length $P=kB$, generation starts at $b_0=P/B$. With $K=\widetilde{L}/B$ and $\mathcal{C}_b=(\tilde{\boldsymbol{x}}^{(<b)},\boldsymbol{H}_b,\boldsymbol{c}_{\mathrm{sem}},\boldsymbol{s})$, the overall conditional distribution $p_{\Theta}$ is:
\begin{equation}
\vspace{-0.5em}
p_{\Theta}\big(
\tilde{\boldsymbol{x}}_{\ge P}
\mid \boldsymbol{I},a_{\mathrm{ref}},\boldsymbol{x}_{<P}
\big) = \prod_{b=b_0}^{K-1} p_{\vartheta}^{\mathrm{blk}}\big(
\tilde{\boldsymbol{x}}^{(b)}
\mid \mathcal{C}_b
\big).
\end{equation}

Each block is learned by using masked diffusion. During training, a subset of positions $\mathcal{M}_b \subseteq [bB, (b+1)B)$ in the active block $b$ is replaced by $\langle\mathrm{MASK}\rangle$ using a cosine schedule $r = \cos(\frac{\pi}{2}\mu), \mu \sim \mathcal{U}(0,1)$. The generator $G_{\vartheta}$ is optimized via cross-entropy in the masked subset $\mathcal{M}_b$. 

During inference, each block is initialized as fully masked and is refined iteratively over $R$ rounds. In each step $t$, $G_{\vartheta}$ predicts uncommitted tokens and ranks them by confidence perturbed by annealed Gumbel noise \cite{chang2022maskgit}. Concretely, $G_{\vartheta}$ re-predicts every still-masked position at round $t$ by sampling a candidate $\hat{x}_j$ per position $j$ with choice temperature $\tau_{c}$ and scores it by: 
\begin{equation}
\vspace{-0.5em}
\operatorname{score}_j=\log p_{\vartheta}\bigl(\hat{x}_j\mid z^{(b)},\widetilde x^{(<b)},H_b,c_{\mathrm{sem}},\boldsymbol{s}\bigr)
+\tau_c\Bigl(1-\frac{t}{R}\Bigr)g_j,
\end{equation}
where $g_j\sim\mathrm{Gumbel}(0,1)$. 
At round $t$, the highest-scoring candidates are committed, leaving $\lfloor B\cos(\pi t/(2R))\rfloor$ positions masked. EOS is suppressed for $j < L_{\min}$. After each completed block, the sequence is truncated before its first $\langle\mathrm{EOS}\rangle$; otherwise the generation continues to the limit $L_{\max}$.

\textbf{Loss Function.} To learn monotonicity-guided alignment, we supervise the conditioning states $\boldsymbol{H}=g_{\phi}(\cdot)$ with unit prediction $\mathcal{L}_{\mathrm{unit}}$ and occupancy prediction objectives $\mathcal{L}_{\mathrm{occ}}$. Two prediction heads map $\boldsymbol{H}_j$ at position $j$ to unit and occupancy distributions, denoted by $p_{\mathrm{unit}}$ and $p_{\mathrm{occ}}$, respectively. During training, $g_{\mathrm{\phi}}$ produces $L+B$ states in a single forward pass, of which $G_{\vartheta}$ uses the first $\widetilde{L}$ states.  Occupancy supervision covers all $L+B$ states and is used only during training.
\begin{equation}
\begin{aligned}
\mathcal{L}_{\mathrm{unit}}
&=-\frac{1}{L}\sum_{j=P}^{L-1}\log p_{\mathrm{unit}}(x_j\mid \boldsymbol{H}_j),\\
\mathcal{L}_{\mathrm{occ}}
&=-\frac{1}{L+B}\sum_{j=0}^{L+B-1}\log p_{\mathrm{occ}}(\mathbb{I}[j<L]\mid \boldsymbol{H}_j),
\end{aligned}
\end{equation}
where $\mathbb{I}[j<L]$ indicates whether the position $j$ is within the target sequence. We form the MQA training loss as: $\mathcal{L}_{\mathrm{MQA}} = \mathcal{L}_{\mathrm{unit}}+\mathcal{L}_{\mathrm{occ}}$.
The masked prediction objective for the generator $G_{\vartheta}$ can be formulated as:
\begin{equation}
    \mathcal{L}_{\mathrm{mask}} = \frac{
\mathbb{E}_{x, b, r}\!\left[\sum_{j\in\mathcal{M}_b}-\log p_{\vartheta}(\widetilde{x}_j|\boldsymbol{z}^{(b)},\widetilde{\boldsymbol{x}}^{(<b)}, H_b, \boldsymbol{c}_{\mathrm{sem}}, \boldsymbol{s})\right]
}{\mathbb E_{x, b, r}[\lvert\mathcal{M}_b\rvert]}.
\end{equation}
We combine this masked prediction objective with the unit and occupancy losses to obtain the total loss of MGSA: $
    \mathcal{L}_{\mathrm{MGSA}}
    =\mathcal{L}_{\mathrm{mask}}
    +\mathcal{L}_{\mathrm{unit}} + \mathcal{L}_{\mathrm{occ}}.
$

\section{Model Architecture}
\label{sec:model_arch}
\textbf{Semantic Encoder.} A frozen BLIP2 Q-former first extracts 32 visual tokens from the image $\boldsymbol{I}$ \cite{li2023blip2}, providing high-level visual features for semantic alignment. The SE $\boldsymbol{f}_{\gamma}$, parameterized as a 12-layer base-sized Transformer decoder with 320 learnable queries, maps them to semantic memory $\boldsymbol{M} \in \mathbb{R}^{320 \times 768}$. RoPE is used for positional encoding \cite{su2024roformer}.

\textbf{Speaker conditioning.} Reference speech $a_{\mathrm{ref}}$ provides speaker conditioning $\boldsymbol{s}$ defined in Sec.\,\ref{sec:prob_form}. $a_{\mathrm{ref}}$ is encoded into an ECAPA-TDNN speaker vector \cite{desplanques20_interspeech} and four FACodec timbre and prosody vectors \cite{ju2024naturalspeech3}, following the practice of \cite{yang2025simplespeech2}. Each vector is $\ell_2$-normalized to eliminate magnitude, similar to \cite{du2024cosyvoice2}. The SE concatenates $\boldsymbol{s}$ with visual features along the feature dimension, while MQA appends the projected speaker token $W_s\boldsymbol{s}$ to $\boldsymbol{M}$ along the sequence dimension.

\textbf{Monotonic Query Aligner.} MQA $\boldsymbol{g_{\phi}}$ is a 2-layer base-sized Transformer decoder with a learned query $\boldsymbol{q}$ replicated to the requested sequence length. We use RoPE \cite{su2024roformer} to assign unit positions. We formulate the attention mechanism of $g_{\phi}$ using a block-causal self-attention mask $\boldsymbol{\mathcal{A}}^{\mathrm{slf}}_{j, k}\in\{0, 1\}$ \cite{arriola2025block} for query and key positions $(j,k)$ and a soft monotonic bias $\boldsymbol{\mathcal{B}}^{\mathrm{crs}}_{j,f}$ for cross-attention logits between unit position $j$ and memory slot $f$:
\begin{equation}
\vspace{-0.5em}
\begin{aligned}
    & \boldsymbol{\mathcal{A}}^{\mathrm{slf}}_{j,k} = \mathbb{I}[\lfloor k/B \rfloor \le \lfloor j/B\rfloor], \\
    & \boldsymbol{\mathcal{B}}^{\mathrm{crs}}_{j,f} = - \alpha|f-c(j)|, \ 0 \leq f < F,
\end{aligned}
\end{equation}
where $\alpha=\operatorname{softplus}(\tilde{\alpha}) > 0$ and $c(j) = \min(j/\rho, F-1)$. $\rho$ is the nominal number of units per memory slot. The appended speaker token $W_s\boldsymbol{s}$ has zero bias, i.e., $\mathcal{B}^{\mathrm{crs}}_{j,F}=0$. $\boldsymbol{\mathcal{A}}^{\mathrm{slf}}$ restricts each query to its own and preceding blocks. $\boldsymbol{\mathcal{B}}^{\mathrm{crs}}$ encourages attention near $c(j)$ with a learnable bias strength $\alpha$. 

\textbf{Block Masked Diffusion Generator.} The generator $G_{\vartheta}$ is an 8-layer, 512-dimensional Diffusion Transformer (DiT) \cite{peebles2023dit} with 8 attention heads and local Canon layers \cite{Allen2025-canon}. The input $\boldsymbol{u}_j$ to $G_{\vartheta}$ combines the token embedding $\boldsymbol{E}_{\mathrm{tok}}$ with $\boldsymbol{H}_{b,j}$ that is linearly projected to 512 dimensions by $W_H$:
\begin{equation}
    \boldsymbol{u}_j=\boldsymbol{E}_{\mathrm{tok}}(z_j)+W_H\boldsymbol{H}_{b,j} + \boldsymbol{b}_H,
\end{equation}
where $\boldsymbol{b}_H$ is a learnable bias and $z_j$ is the unit, mask, or EOS token at position $j$. We use adaptive layer-normalization to inject $\boldsymbol{c}_{\mathrm{sem}}$ and $\boldsymbol{s}$ \cite{peebles2023dit}. Classifier-free guidance (CFG) is applied only to image conditioning \cite{ho2021cfg}. The null branch replaces visual features with zeros while retaining $\boldsymbol{s}$.

\begin{table}[t!]
    \centering
    \small
    \setlength{\tabcolsep}{5pt}
    \renewcommand{\arraystretch}{0.95}
    \begin{tabular}{lccccccccc}
    \specialrule{.2em}{.1em}{.1em}
    \multicolumn{2}{l}{Setting}
    & B1 $\uparrow$ &B4 $\uparrow$ & M $\uparrow$
    & R $\uparrow$ & C $\uparrow$ \\
    \specialrule{.1em}{.05em}{.05em}

    
    \multicolumn{4}{l}{\textit{Alignment \& Visual Conditioning}} \\
    \multicolumn{2}{l}{MGSA (w/o MQA)}
    & \underline{56.4} & \underline{15.7} & \underline{17.7} & \underline{42.5} & \underline{33.3} \\
    \multicolumn{2}{l}{MGSA (mismatched img.)}
    & 34.5 & 2.71 & 8.08 & 25.4 & 3.61 \\
    \midrule
    \multicolumn{6}{l}{\textit{Block Diffusion}} \\
    \multicolumn{2}{l}{AR}
    & 38.2 & 6.9 & 17.2 & 34.4 & 15.0 \\
    \multicolumn{2}{l}{AR w/ mismatched img.}
    & 23.9 & 1.07 & 8.19 & 21.4 & 1.55 \\
    \midrule
    \multicolumn{2}{l}{MGSA (full)}
    & $\mathbf{61.3}$ & $\mathbf{18.4}$ & $\mathbf{20.0}$ & $\mathbf{45.6}$ & $\mathbf{41.5}$ \\
    \specialrule{.2em}{.1em}{.1em}
    \end{tabular}
    \vspace{-0.5em}
    \caption{Ablations at $200k$ steps via the \emph{cross-sentence} protocol. \textbf{Bold} and \underline{underline} denotes the best and the second best.}
    \label{tab:ablation}
    \vspace{-1.0em}
\end{table}

\begin{table}[t]
\centering
\small
\setlength{\tabcolsep}{2pt}
\renewcommand{\arraystretch}{0.95}
\begin{tabular*}{\columnwidth}
{@{\extracolsep{\fill}}lcccccc@{}}
\specialrule{.2em}{.1em}{.1em}
Method & \shortstack{Speaker\\Setting}
& B1 $\uparrow$ & B4 $\uparrow$ & M $\uparrow$ & R $\uparrow$ & C $\uparrow$ \\
\specialrule{.1em}{.05em}{.05em}

SAT \cite{hsu2021sat}
& Single & -- & 11.6 & 14.1 & 39.0 & 23.2 \\
SAT-FT \cite{hsu2021sat}
& Single & -- & 12.6 & 14.5 & 39.1 & 24.2 \\
E-I2S \cite{effendi2021end}
& Single & -- & 14.8 & 17.4 & 45.8 & 32.9 \\
Im2Sp \cite{kim2024towards} (CLIP init.)
& Single & -- & 17.7 & 20.6 & 45.9 & 45.8 \\
Im2Sp \cite{kim2024towards} (GiT init.)
& Single & -- & \textbf{20.6} & \textbf{22.0} & \textbf{48.4} & \textbf{53.6} \\

SAS \cite{wang2021show}
& Single & 29.6 & 3.5 & 11.3 & 23.2 & 8.0 \\
PixVoxLM \cite{f2p2025} (Delay)
& Single & 48.1 & 11.5 & 15.2 & 35.8 & 25.5 \\
\textbf{MGSA} (\emph{same-sent.})
& \textbf{Multiple} & 62.4 & \underline{19.2} & 20.8 & 46.3 & 45.0 \\
\textbf{MGSA} (\emph{cross-sent.})
& \textbf{Multiple} & 62.3 & 19.0 & 20.6 & 46.1 & 44.9 \\
\textbf{MGSA} (\emph{unseen})
& \textbf{Multiple} & \textbf{63.8} & 18.4 & \underline{21.2} & \underline{47.2} & \underline{46.2} \\
\specialrule{.1em}{.05em}{.05em}

Setting
& \multicolumn{2}{c}{DistillMOS $\uparrow$}
& \multicolumn{2}{c}{UTMOS $\uparrow$}
& \multicolumn{2}{c}{SIM $\uparrow$} \\
\cmidrule(lr){2-3}\cmidrule(lr){4-5}\cmidrule(l){6-7}
Ground truth
& \multicolumn{2}{c}{$3.65 \pm 0.58$}
& \multicolumn{2}{c}{$3.50 \pm 0.68$}
& \multicolumn{2}{c}{$0.973$} \\
\midrule
MGSA (\emph{same-sent.})
& \multicolumn{2}{c}{$\underline{3.66 \pm 0.59}$}
& \multicolumn{2}{c}{$\underline{3.99 \pm 0.47}$}
& \multicolumn{2}{c}{$\mathbf{0.940}$} \\
MGSA (\emph{cross-sent.})
& \multicolumn{2}{c}{$\underline{3.66 \pm 0.59}$}
& \multicolumn{2}{c}{$3.95 \pm 0.47$}
& \multicolumn{2}{c}{$\underline{0.935}$} \\
MGSA (\emph{unseen})
& \multicolumn{2}{c}{$\mathbf{\mathbf{3.98} \pm \mathbf{0.43}}$}
& \multicolumn{2}{c}{$\mathbf{\mathbf{4.44} \pm \mathbf{0.08}}$}
& \multicolumn{2}{c}{$\underline{0.935}$} \\
\specialrule{.2em}{.1em}{.1em}
\end{tabular*}
\vspace{-1.0em}
\caption{Baseline comparisons and speech quality evaluations. All evaluations use $P=0$. $\pm$: standard deviation.}
\label{tab:baselines}
\label{tab:naturalness}
\vspace{-1.5em}
\end{table}

\vspace{-0.2em}
\section{Experiments}
\label{sec:exp}
\subsection{Experimental Setup}
\vspace{-0.3em}
\label{ssec:exp_setup}
\textbf{Data Preparation.} We use Flickr8k-Audio \cite{harwath2015deep} and SpokenCOCO \cite{hsu2021sat} together  under Karpathy split \cite{karpathy2015deep} for training and test, following prior works \cite{hsu2021sat, wang2021show, effendi2021end, kim2024towards, f2p2025}. We apply LibriTTS-R test reference of 38 speakers for unseen speaker evaluation \cite{koizumi2023librittsr}. We extract semantic speech units using Cosyvoice 2's automatic-speech-recognition-supervised $\mathcal{S}^3$ tokenizer and use its acoustic decoder for U2S synthesis \cite{du2024cosyvoice2}.

\textbf{Training Strategy.} Both MQA $g_{\phi}$ and generator $G_{\vartheta}$ operate with a block size $B = 16$ ($0.64~\mathrm{s}$ at $25~\mathrm{Hz}$), following \cite{seo2026chatterboxflash}. During training, speaker conditioning $\boldsymbol{s}$ is extracted from an utterance sampled uniformly from the target speaker's utterance pool. The model is optimized using scheduler-free AdamW \cite{defazio2024schedulefree} with a batch size of 128, a learning rate of $5e^{-4}$, and $300k$ total steps with a $5\mathrm{k}$-step warmup. We set $L_{\min}=0$, $L_{\max}=400$, and $\rho=2$ to cover the maximum length of the training dataset. We set $p_{\mathrm{cfg}} = 0.1$ for CFG training and the CFG scale to 1.5 for inference. During inference, we set $R=64$ and $\tau_{c}=0.0$.

\textbf{Evaluation Metrics.} Following previous works, we apply an off-the-shelf ASR model (wav2vec2-Large finetuned on LibriSpeech 960h) to transcribe synthesized speech and evaluate BLEU (B1 \& B4)\,\cite{papineni2002bleu}, METEOR (M)\,\cite{banerjee2005meteor}, ROUGE-L (R)\,\cite{lin2004rouge}, and CIDEr (C)\,\cite{vedantam2015cider}. We assess naturalness using DistillMOS \cite{stahl2025distillmos} and UTMOS \cite{saeki2022utmos}. Speaker similarity (SIM) is evaluated using WavLM-SV following the practice of \cite{wang2023neural}. 

\textbf{Reference Protocols.} Inspired by \cite{lee2025dittotts, yang2025simplespeech2, wang2023neural}, we evaluate three reference speech conditions as: (1) the target utterance itself (\emph{same-sentence}); (2) a different utterance from the same speaker (\emph{cross-sentence}); (3) and an external utterance from an unseen speaker during MGSA training (\emph{unseen}). All 176 Flickr8k-Audio test speakers appear in MGSA training. 

\textbf{Baselines.} We compare with SAT \cite{hsu2021sat}, E-I2S \cite{effendi2021end}, SAS \cite{wang2021show}, Im2Sp \cite{kim2024towards}, and PixVoxLM \cite{f2p2025} on the Flickr8k-Audio \cite{harwath2015deep} test split using published scores. All reported baselines use single-speaker synthesis. SAS predicts mel spectrograms, while SAT and E-I2S use discrete speech units from separately trained VQ-VAE-based U2S models. Im2Sp trains an AR-based I2U model using 200-cluster HuBERT units and a unit-based vocoder. PixVoxLM predicts EnCodec tokens with fine-tuning BLIP \cite{li2022blip}. In contrast, MGSA supports \emph{zero-shot multispeaker} synthesis conditioned on reference speech from unseen speakers. Remarkably, although captioning metrics remain directly comparable, baselines benefit from a more constrained single-speaker setting and do not face an additional generalization challenge to unseen speakers.

\begin{table}[t]
\centering
\small
\setlength{\tabcolsep}{4pt}
\begin{tabular*}{\columnwidth}
{@{\extracolsep{\fill}}lc@{}}
\specialrule{.2em}{.1em}{.1em}
Speech pair & SIM \\
\midrule
Generated speech / unseen reference & $0.935 \pm 0.043$ \\
Natural recording / same speaker         & $0.946 \pm 0.040$ \\
Natural recording / different speakers   & $0.658 \pm 0.162$ \\
\specialrule{.2em}{.1em}{.1em}
\end{tabular*}
\vspace{-0.5em}
\caption{Speaker similarity evaluation for unseen speaker synthesis, compared with same- and different-speaker pairs of natural LibriTTS-R recordings.}
\label{tab:unseen_sim}
\vspace{-1.5em}
\end{table}

\vspace{-0.8em}
\subsection{Experimental Results}
\label{ssec:exp_results}
\vspace{-0.2em}
\textbf{Alignment and Visual Conditioning.} We remove MQA $g_{\phi}$ while retaining $\boldsymbol{c}_{\mathrm{sem}}$ and $\boldsymbol{s}$ to validate its effectiveness and separately test mismatched images to evaluate visual conditioning. Results in Table\,\ref{tab:ablation} support the effectiveness of MQA and confirm that MGSA uses image conditioning.

\textbf{Block Diffusion.} We replace the blockwise masked diffusion with an AR generator and repeat the mismatched-image conditioning test. Table\,\ref{tab:ablation} shows that the AR variant performs worse and further degrades with mismatched images, supporting the benefit of blockwise masked diffusion.

\textbf{Unseen Speaker.} As shown in Table\,\ref{tab:baselines}, with reference speech from unseen LibriTTS-R speakers, MGSA achieves captioning performance comparable to that in the seen-speaker settings and obtains higher DistillMOS and UTMOS scores. Its generated-to-reference SIM is close to that of natural LibriTTS-R recordings from the same speakers and well above that of different speakers. Results in Tables\,\ref{tab:baselines} and \ref{tab:unseen_sim} support MGSA's capability on zero-shot synthesis while maintaining good captioning performance.

\textbf{Comparison against single-speaker baselines.} Results in Table\,\ref{tab:baselines} show that MGSA outperforms SAS, SAT, SAT-FT, and PixVoxLM. Under the \emph{unseen} protocol, it also exceeds the Im2Sp variant with a CLIP-initialized encoder and a randomly initialized decoder. The GiT-initialized Im2Sp variant achieves higher captioning scores, while MGSA allows zero-shot multispeaker synthesis for unseen speakers. 


\section{Conclusion \& Limitation} 
\label{sec:conclu}
We present MGSA, a new framework for zero-shot multispeaker synthesis that combines monotonicity-guided alignment and block diffusion. Experiments show that MGSA achieves competitive performance against single-speaker baselines while supporting zero-shot multispeaker synthesis. Ablation studies indicate the effectiveness of the alignment mechanism and blockwise diffusion designs. However, captioning performance remains a limitation for practical application, motivating further improvement in visual grounding and content accuracy in the I2U model.

\newpage
\bibliographystyle{IEEEbib}
\let\oldthebibliography\thebibliography
\let\endoldthebibliography\endthebibliography

\renewenvironment{thebibliography}[1]{
  \oldthebibliography{#1}
  \setlength{\itemsep}{1pt}
  \setlength{\parskip}{3pt}
  \footnotesize
}{
  \endoldthebibliography
}
\bibliography{refs}

\end{document}